\documentclass[10pt, a4paper, 3p]{elsarticle}
\makeatletter
\def\ps@pprintTitle{%
 \let\@oddhead\@empty
 \let\@evenhead\@empty
 \def\@oddfoot{\centerline{\thepage}}%
 \let\@evenfoot\@oddfoot}
\makeatother
\usepackage[utf8]{inputenc}
\usepackage{amsmath}
\usepackage{amsfonts}
\usepackage{amssymb}
\usepackage{mathrsfs}
\usepackage{color}
\usepackage{slashed}

\def\be{\begin{equation}}
\def\ee{\end{equation}\noindent}
\def\bear{\begin{eqnarray}}
\def\ear{\end{eqnarray}\noindent}

\def\ddel{{}^\bullet\! \Delta}

\def\ddeld{{}^{\bullet}\! \Delta^{\hskip -.5mm \bullet}}

\def\slash#1{#1\!\!\!\raise.15ex\hbox {/}}
\newcommand{\slD}{\,\raise.15ex\hbox{$/$}\kern-.27em\hbox{$\!\!\!D$}}
\newcommand{\slpartial}{\raise.15ex\hbox{$/$}\kern-.57em\hbox{$\partial$}}

\newcommand{\e}{\textrm{e}}

\newcommand{\Det}{\textrm{Det}}

\newcommand{\ep}{\varepsilon}

\newcommand{\linep}{\textrm{lin } \ep_{1}\ldots\ep_{N}}

\newcommand{\GBtij}{G_{B}\left(\tau_{i}, \tau_{j}\right)}

\newcommand{\GBddtij}{\ddot{G}_{B}\left(\tau_{i}, \tau_{j}\right)}
\newcommand{\GBij}{G_{Bij}}

\newcommand{\GBdij}{\dot{G}_{B ij}}

\newcommand{\GBddij}{\ddot{G}_{B_{ij}}}

\newcommand{\GFtij}{G_{F}\left(\tau_{i}, \tau_{j}\right)}

\newcommand{\GFij}{G_{Fij}}

\newcommand{\GBctij}{\mathcal{G}_{B}\left(\tau_{i}, \tau_{j}\right)}

\newcommand{\GFctij}{\mathcal{G}_{F}\left(\tau_{i}, \tau_{j}\right)}

\newcommand{\Zz}{\mathcal{Z}}

\newcommand{\pb}{\textrm{PB}}

\begin{document}

\begin{frontmatter}
\title{Quantum mechanical path integrals in the first quantised approach to quantum field theory}	

\author[a]{James P. Edwards\corref{cor1}\fnref{fn1}}
\ead{jedwards@ifm.umich.mx}
\author[a]{Christian Schubert}
\ead{schubert@ifm.umich.mx}

\address[a]{Instituto de F\'{\i}sica y Matem\'aticas,\\
Universidad Michoacana de San Nicol\'as de Hidalgo,\\
Edificio C-3, Apdo. Postal 2-82, C.P. 58040, Morelia, Michoac\'an, M\'exico\\}
\cortext[cor1]{Corresponding author.}
%\fntext[fn1]{\textit{Telephone:} +52 443 322 3500: 4140}

\begin{abstract}
Perturbative quantum field theory usually uses second quantisation and Feynman diagrams. 
The worldline formalism provides an alternative approach based on first quantised particle path integrals, similar
in spirit to string perturbation theory. Here we review the history, main features and present applications of the formalism. Our emphasis is on recent developments such as the path integral representation of open
fermion lines, the description of colour using auxiliary worldline fields, incorporation of higher spin, and  extension of the formalism to
non-commutative space. 
\end{abstract}

\end{frontmatter}
\section{Introduction}
Path integrals in quantum field theory usually refer to functional integrals over field configurations as in the second quantised approach. This approach -- originally going back to Feynman \cite{Feynman:1948ur}, 
building on work by Dirac -- came to dominate the development of quantum electrodynamics and later theories such as quantum chromodynamics and gravity. 
%Scalar and spinor fields can be described in this framework and scattering amplitudes follow via the LSZ identification \cite{LSZ1, LSZ2} of S-matrix elements as the residue of the poles in on-shell field correlators.
This is despite the fact that in the early fifties Feynman himself showed that the S-matrix of scalar or spinor QED can be represented in terms of first quantised particle path integrals 
interconnected by photon propagators in all possible ways. For the simplest case, the scalar propagator coupled to an external Maxwell field $A^{\mu}(x)$ \cite{FeynEM}, this ``worldline representation'' takes the form:

\begin{equation}
	D^{x'x}[A] = \int_{0}^{\infty} dT \, \e^{-m^{2} T} \int_{x(0) = x}^{x(T) = x'} \hspace{-1em}\mathscr{D}x \, \e^{- \int_{0}^{T} d\tau\, \left[ \frac{\dot{x}^{2}}{4} + ie\dot{x} \cdot A(x(\tau)) \right]}.
	\label{DScalarPI}
\end{equation}
Here $x$ and $x'$ are fixed points in space-time, and the path integral $\mathscr{D}x$ runs over all possible trajectories of the scalar particle that lead from $x$ to $x'$ in the proper-time $T$, to be integrated over in the
end. Large proper-times are suppressed by the factor $\e^{-m^2T}$, with $m$ the mass of the scalar particle
%\footnote{On a fundamental level, the integral $\int_0^{\infty}dT$ can be seen as a modular integral left-over from the gauge-fixing of the group of diffeomorphisms on an interval from a reparameterisation invariant from of the point particle action.}.  
%The proper-time $T$ then appears as the only {\it modulus}, i.e. as the only invariant that can be formed from the {\it einbein} field $e(\tau)$, namely as its integral $T = \frac{1}{2} \int_0^1 d\tau e(\tau)$; see below.}. 
However, until the early nineties this approach was considered of theoretical rather than practical interest, and on the few occasions where it was used as a calculational tool \cite{Halpern:1977he,AAM} it was in the non-perturbative
context rather than for standard perturbative computations of amplitudes or cross sections.  

The advent of string theory reignited interest in the first quantised approach, not least because its infinite tension limit can be related to quantum field theory \cite{String1, String2}. The Polyakov path integral, describing a single string rather than the multi-particle states of field theory, was shown to reproduce S-matrix elements of select field theories such as Yang-Mills \cite{GerNev, YMBeta}. Bern and Kosower studied this infinite tension limit in the 1990s \cite{BK2, BK3}, arriving at their ``Master Formulae'' integral representations of one-loop on-shell N-gluon amplitudes, extended to graviton amplitudes by Bern, Dunbar and Shimada \cite{Bern:1993wt,Dunbar:1994bn}.  
One significant advantage of string theory is that fewer diagrams contribute to a given process than in the field theory limits. Moreover, the processes are better organised to display gauge invariance and are free of undetermined loop momenta.

Shortly after these string-based developments, Strassler rederived the Bern-Kosower Master Formula \cite{Strass1} for the gluon scattering case using a non-abelian generalisation of Feynman's QED formalism and performing the
path integral in a way analogous to string theory, as had been suggested earlier by Polyakov \cite{Polyakov:1987ez}.  
This finally led to the recognition that this ``worldline formalism'' has certain advantages, similar to those of first quantised string perturbation theory, and to a concerted effort to extend it to other field theory couplings 
\cite{Yuk,Ax,YukAx1,YukAx2,McK1,McK2}
as well as to quantum gravity \cite{GravGh2,GravTr1,GravTr2,GravSp2}. 
Although a review of the formalism exists \cite{ChrisRev} we update it here by summarising its applications and advantages for computational efficiency as a reorganisation of the physical information of field theory, focusing on recent developments, where quantum mechanical path integrals have been applied to higher spin fields, non-commutative field theory and gravity, numerical calculation of the path integral and more. 

We will present three approaches to the calculation of worldline path integrals: (i) direct numerical computation (ii) semiclassical approximation  (iii) analytic gaussian evaluation (the ``string-inspired approach'').
A variational approach has been championed by Rosenfelder \cite{Alexandrou:2000qk,BarroBergflodt:2004qa} but will not be discussed here. 

\section{Worldline numerics}
We start with the numerical ``worldline Monte Carlo'' approach, a direct numerical calculation of the discretised path integral. Although this method is straightforward, 
and in quantum mechanics has been used to evaluate path integrals for decades (e.g.,  \cite{FeynKac1, FeynKac2, MonteRev, Makri}), surprisingly its application to the relativistic case essentially began only in 1996 with the work of 
Nieuwenhuis and Tjon \cite{Tjon} on scalar bound states.
 
The numerical approach was further developed in \cite{GiesMagnetic, GiesClouds}, where various algorithms for generating a finite ensemble of loops, $\{x_{n}\}_{n = 1}^{N_{L}}$, were proposed. These paths are generated such that the distribution in their velocities corresponds to the kinetic term in the particle action (in euclidean space this gives the Wiener measure on the trajectories).
The line integral of the worldline action is then computed numerically by splitting the path into $N_{p}$ segments (note this is a time discretisation only, maintaining continuous spatial coordinates). 
For example, the path integral describing propagation of a scalar particle with self-interaction potential $U(\Phi(x))$ from $x$ to $x'$ in spacetime
in the fixed proper-time $T$ becomes, after the discretisation,  

\begin{align}
	\int_{x(0) = x}^{x(T) = x'}\hspace{-1em} \mathscr{D}x \, e^{-\int_{0}^{T} dt\left[ \frac{\dot{x}^{2}}{4} + V(x(t))\right]}  \longrightarrow \frac{e^{-\frac{(x - x')^{2}}{4T}}}{(4 \pi T)^{\frac{D}{2}}}N_{L}^{-1} \sum_{n = 1}^{N_{L}} e^{- \frac{T}{N_{p}} \sum_{i = 1}^{N_{p}} V\left(x_{n}(\frac{i}{N_{p}} ) \right) }.
\end{align}
Here the potential entering the worldline Lagrangian is just $V(x) \equiv U''(\Phi(x))$. We will work in euclidean spacetime
throughout this review. For efficiency one may reuse unit loops by scaling $t \rightarrow Tu$, expanding $x(\tau) = x_{0} + q(\tau)$ and using the field-redefinition $q(t) \rightarrow \sqrt{T}q(t) = \sqrt{T}q(Tu)$. For the ensemble average $N_{L}$ must be chosen sufficiently large that a good sampling of the space of trajectories results. 

Worldline numerics have been applied to estimate quantum effective actions \cite{gisava} and study vacuum polarisation and Schwinger pair creation in non-homogeneous background fields \cite{holg3, NumbPol}, with results consistent with analytic perturbation theory. 
They have been remarkably successful in the computation of the Casimir energy with Dirichlet boundary conditions, allowing a treatment of practically arbitrary surfaces \cite{GiesCasimir} through a study of the energy momentum tensor at given spatial points \cite{EMCas1}, including curved geometries \cite{CasCurve1}. 
%Such work would be challenging indeed based upon an expansion of the worldline action, but the geometry of the surface is easily modified in a numerical approach. 
An extension of the numerical approach to include fermionic degrees of freedom can be found in \cite{NumF1}. 

%Finally we comment that lately some studies have adapted the worldline numerics approach to the numerical investigation of non-relativistic systems. This has provided an alternative numerical treatment of the quantum mechanical kernel and estimation of ground state energies and wave functions \cite{TesisAn, UsReg}, where a new algorithm for the generation of ensmbles of worldlines (open or closed) was also included. One drawback of the numeric approach is that it can be applied only in the case that the theory admits continuation to Euclidean space. Furthermore, in both the relativistic and quantum mechanical case, application of worldline numerics has found it crucial to get as good a sampling as possible, given computational constraints, of the potential. This has led to the numerical fits to numerical samplings of the line integrals of the potentials \cite{gisava} and motivated theoretic study of these distributions and the contribution to the path integral of trajectories passing through fixed points in space \cite{HzPv}. 

\section{Worldline instantons and Schwinger pair creation}
The second approach is the semiclassical approximation. For path integrals we recall it means that the whole space of trajectories (or fields)
be replaced by just a single configuration solving the equations of motion, such that the action is stationary on it. At leading order the path integral is approximated by the exponential of the negative action on that classical trajectory, $\e^{-S[x_{\rm cl}]}$. Due to exponential fall-off this is often
sufficient for order-of-magnitude estimates. The prefactor of the exponential involves the determinant of the Hessian of fluctuations around the classical trajectory that is usually much harder to estimate.

In an early brilliant application of the worldline representation, Affleck et al. \cite{AAM} in 1982 applied the semiclassical approximation to the important problem
of Sauter-Schwinger pair production from vacuum by an electric field. The rate for this process, which was predicted by Sauter in 1931, is equal to
twice the imaginary part of the effective action in the field, $\Gamma[E]$, by virtue of the optical theorem. In 1951, Schwinger \cite{Schwinger} calculated this quantity for a constant field at the one-loop
level,  finding the following decomposition in terms of ``Schwinger exponentials'':
\bear
{\rm Im} {\cal L}(E) &=&  \frac{m^4}{8\pi^3}
\Bigl(\frac{eE}{m^2}\Bigr)^2\, \sum_{n=1}^\infty \frac{1}{n^2}
\,\exp\left[- n \pi \frac{m^2}{eE}\right] \, .
\label{schwinger}
\ear
Thus pair creation is always possible, but exponentially suppressed below the critical field strength 
\bear
E_{\rm cr} = m^2/e \approx 10^8 V/m \, .
\label{st}
\ear
In \cite{AAM} this formula (or rather its analogue for scalar QED, which differs from \eqref{schwinger} only by a factor of $\frac{1}{2}$ and sign changes)
was rederived applying a semiclassical approximation to the worldline representation of the effective action. In the worldline formalism, the (one-loop)
effective action is given by 

\begin{equation}
	\Gamma[A] = \int_{0}^{\infty} \frac{dT}{T} e^{-m^{2}T} \oint \mathscr{D}x \, \e^{-\int_{0}^{T} d\tau \left[ \frac{\dot{x}^{2}}{4} + ie \dot{x}  \cdot A(x(\tau)) \right]} \, .
	\label{GamScalarPI}
\end{equation}
The difference to the propagator case, \eqref{DScalarPI}, is that the path integration now runs over the space of closed trajectories with fixed periodicity $T$.
The equation of motion is simply the Lorentz force equation. 
%$\ddot x^{\mu} = 2ie F^{\mu\nu}\dot x_{\nu}$.
In the constant electric field case, the motion in euclidean space is circular (rather than hyperbolic
as in Minkowski space), with a curvature proportional to the field strength. Choosing the field to point in the $z$ direction, the solutions take the form
(we have Wick-rotated $t\to -ix_4$)
\bear
x(\tau) =\bigl (x_1,x_2,{\cal N}\cos (2eE\tau),{\cal N}\sin(2eE\tau)\bigr) \, .
\label{defwlinst}
\ear
Here $x_1,x_2$ and ${\cal N}>0$ are arbitrary constants. They fulfill the periodicity condition when the proper-time takes one of the values
$T_n = \frac{n\pi}{eE}; \quad n = 1,2,\ldots $. 
These periodic classical solutions, called ``worldline instantons,'' lead to poles in the proper-time integral at the locations $T_{n}$,
and the $n$-th pole generates the corresponding term in Schwinger's formula for the imaginary part \eqref{schwinger}. Moreover, the fluctuation determinant produces the correct prefactors in that equation \cite{AAM}.

Much later, in 2005, this worldline instanton method was generalised in \cite{63} to the spinor QED case, as well as to non-constant fields. 
In some cases the instantons can be obtained in closed form, in others they are constructed by a numerical solution of the Lorentz force equation \cite{Dunne:2006ur}.
%In 'cite{64} an extremely convenient method was found for the computation of the fluctuation determinant around a general worldline instanton.
In recent years the worldline instanton method has become widely used for the computation of Schwinger pair creation rates particularly in laser physics. 
There the search is on for laser configurations that would  bring down the Schwinger threshhold \eqref{st},
and the method has the great advantage that one can not only obtain an estimate of the rate but also information on its variation with the external parameters, such as switching on an additional laser field; 
see \cite{Schneider:2014mla}.  
It is also very useful for studying  the interference effects present when the temporal profile of a laser has subcycle structure \cite{PCT2}.
Recent extensions of the method include pair creation in de Sitter space \cite{Guts:2013dha} and Hawking radiation \cite{Dumlu:2017kfp} (see below for the worldline approach in a gravitational background).

The founding paper by \cite{AAM} contains also an interesting, though non-rigorous, attempt to use the one-loop instanton \eqref{defwlinst} for a generalisation of the Schwinger formula \eqref{schwinger} to arbitrary loop order. This is of great potential interest for the study of the structure of QED perturbation theory, see \cite{Huet:2017ydx} for a discussion.  

\section{Scalar QED}
The remainder of our review will be devoted to the third and most widely used method of calculation of worldline path integrals, called ``string-inspired'' on 
account of its genesis as a spin-off of developments in string perturbation theory. Here all path integrals are reduced to gaussian ones, which can be computed using Green's functions and functional
determinants in the one-dimensional worldline theory. 
%This reduction usually requires a perturbation expansion around the case of the free particle, or some other exactly solvable case.
Despite providing a fairly universal approach to QFT, so far it has mostly found applications in gauge theory. 
This is partly because the worldline approach allows one to take advantage of gauge symmetries in ways that would not be
easy to emulate using Feynman diagrams. We will start with scalar QED at the tree level.

\subsection{Tree level amplitudes}

Thus we return to Feynman's worldline representation of the propagator in a field, \eqref{DScalarPI}.
In the string-inspired approach, a gaussian form for the path integral is achieved by expanding the ``interaction exponential'' 
$\e^{- ie\int_{0}^{T} d\tau\,  \dot{x} \cdot A(x(\tau))}$, and then either (i) expanding $A(x)$ around some fixed point in space-time or
(ii) specialising it to plane waves of fixed momenta and polarisations, $A_{\mu}(x) \rightarrow \sum_{i = 1}^{N}\ep_{i\mu}\, \e^{i k_{i} \cdot x}$.  
In the first case, we obtain the ``heat-kernel expansion'' or ``higher-derivative expansion'' of the propagator (if the expansion is organised according
to mass dimension or number of derivatives involved respectively). The second case yields the amplitude for the scalar to emit or absorb, during its propagation, 
$N$ photons of prescribed momenta and polarisations. Each photon is represented in the path integral by a ``photon vertex operator''  
\begin{eqnarray}
V_{\gamma}[k, \ep] \equiv \int_{0}^{T}d\tau\, \varepsilon \cdot \dot{x} \,\e^{i k \cdot x} = \int_{0}^{T}d\tau \,\e^{i k \cdot x +  \varepsilon \cdot \dot{x}} \Big\vert_{\varepsilon}
\label{defvertop}
\end{eqnarray}
that gets integrated along the trajectory of the scalar particle. Note that the integrand turns into a total derivative if one replaces $\varepsilon$ by $k$; 
this gives already a hint that, in this formalism, there is an intimate relation between gauge invariance and integration by parts. 

\begin{figure}[h]
  \centering
   \includegraphics[width=1.1\textwidth]{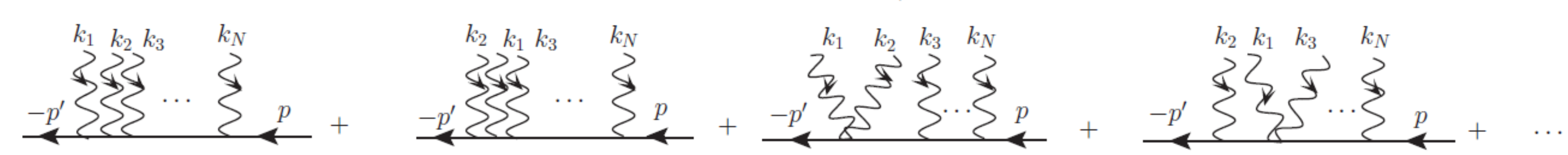}
\caption{Multi-photon Compton scattering diagram.} 
\label{fig-multiphoton}
\end{figure}

In either case, the computation of the final gaussian path integrals requires, apart from the global normalisation which can be fixed by
comparison with the known free propagator, only knowledge of 
%the inverse of the free kinetic term in the path integral, i.e.
the Green's function of the second derivative operator $-\frac{1}{4}\frac{d^2}{d\tau ^2}$. It is convenient to perform the path integration 
%not directly in the trajectory $x(\tau)$,but rather 
in the fluctuation $q(\tau)$ of the trajectory $x(\tau)$ around the straight-line path $x_0(\tau) = x + \frac{\tau}{T} (x'-x)$ from $x$ to $x'$. 
This reduces the boundary conditions to Dirichlet ones, $q(0) = q(T) =0$, and the Green's function is 
\begin{equation}
	\Delta(\tau_{i}, \tau_{j}) = \frac{\left|\tau_{i} - \tau_{j}\right|}{2} - \frac{\tau_{i} + \tau_{j}}{2}  + \frac{\tau_{i}\tau_{j}}{T}.
\end{equation}
Using the exponentiated form of the photon vertex operator given above in \eqref{defvertop} the gaussian integration can be done in closed form for any $N$
by completing-the-square.
Fourier transforming also the scalar line to momentum space, one obtains the following ``master formula'' \cite{Daik,line2}:
\bear 
D^{p'p}(k_1,\varepsilon_1;\cdots; k_N,\varepsilon_N)&=&(-ie)^N
%(2\pi)^D\delta^D\Big(p+p'+\sum_{i=1}^Nk_i\Big)
\int_0^\infty dT\,{\rm e}^{-m^2T}
%\nonumber\\ &&\hspace{-1.5cm}\times
\prod_{i=1}^N\int_0^Td\tau_i\,{\rm e}^{-Tb^2+\sum_{i,j=1}^N[\Delta_{ij}k_i\cdot k_j-2i\ddel_{ij}\varepsilon_i\cdot k_j-\ddeld_{ij}\varepsilon_i\cdot \varepsilon_j]}\Big\vert_{\varepsilon_1\varepsilon_2\cdots\varepsilon_N}
\, .
\nonumber\\
\label{linemaster} 
\ear
Here we have abbreviated $b \equiv p'+\frac{1}{T}\sum_{i=1}^N(k_i\tau_i-i\varepsilon_i)$
and we use left and right dots on $\Delta(\tau_i,\tau_j)$ to indicate derivatives with respect to the first and the second argument respectively \cite{BastBook}. 
The notation $\Big\vert_{\varepsilon_1\varepsilon_2\cdots\varepsilon_N}$ means that one should expand the exponential and keep only terms containing each polarisation vector once. 
This generating function yields the sum over all orderings of external photon legs, as shown in Fig. \ref{fig-multiphoton}.

\noindent These diagrams represent multi-photon Compton scattering in scalar QED, as well as processes related by crossing. However, we have not yet imposed on-shell conditions, so the integral representation of  the master
formula for the ``photon-dressed'' scalar propagator can also be used to construct higher-loop amplitudes.

\subsection{One-loop N-photon amplitudes}

As for the propagator, from the closed-loop path integral (\ref{GamScalarPI}) we obtain one-loop $N$-photon amplitudes 
by specialising to a plane-wave background, expanding the interaction exponential to $\mathcal{O}(N)$, and retaining the terms containing each $\ep_{i}$ once. 
The only difference is the boundary conditions: a closed loop is written as $x(\tau) = x_0 + q(\tau)$, where $x_0=\frac{1}{T}\int_0^Td\tau\, x(\tau)$ is the loop center of mass,
so that $q(\tau)$ fulfills the condition $\int_0^Td\tau \,q(\tau) = 0$ instead of Dirichlet boundary conditions. The appropriate worldline Green's function is 
\begin{equation}
	\GBtij =  \left|\tau_{i} - \tau_{j}\right| - \frac{(\tau_{i} - \tau_{j})^{2}}{T} - \frac{T}{6} \, .
	\label{GB}
\end{equation}
Note that, contrary to $\Delta(\tau_i,\tau_j)$ above, this Green's function is a function of $\tau_{i} - \tau_{j}$, since translation invariance in proper-time is not broken by the
boundary conditions. Thus here the convention is that `dot' means a derivative with respect to the first variable. 
%It should also be remarked that momentum conservation allows the constant $- \frac{T}{6}$ to be safely omitted in flat-space calculations without affecting physical results. 

Performing the gaussian integration over $q(\tau)$ as before, we obtain, with little effort, the Bern-Kosower Master Formula 
($x_0$ must also be integrated over, and provides the usual momentum conserving $\delta$ function):
\begin{align}
	\Gamma[k_{1}, \ep_{1}; \ldots; k_{N}, \ep_{N}] &= (2\pi)^{D}\delta^{D}\left(\sum_{i = 1}^{N} k_{i}\right)  (-ie)^{N}\int_{0}^{\infty}\frac{dT}{T} (4\pi T)^{-\frac{D}{2}} \e^{-m^{2}T} \nonumber \\
	&\qquad \qquad \times\prod_{i = 1}^{N}\int_{0}^{T} d\tau_{i} \,\e^{\frac{1}{2}\sum_{i, j = 1}^{N} \left[ k_{i} \cdot k_{j} \GBij - 2i k_{i} \cdot \ep_{j} \GBdij + \ep_{i}\cdot \ep_{j} \GBddij \right]}\bigg|_{\linep}.
	\label{GamScalar}
\end{align}
This formula generates parameter integrals for all the (on-shell and off-shell) one-loop $N$-photon amplitudes in scalar QED, written in terms of the Green's function $G_B$ and its first and second derivatives.
%$\GBdij = {\rm sign}(\tau_i-\tau_j) - 2\frac{\tau_i-\tau_j)}{T}$ and $\GBddtij = 2\delta(\tau_{i} - \tau_{j}) - \frac{2}{T}$. 
The $\delta$ - function contained in the latter
produces the seagull diagrams of scalar QED. Although these integrals are directly related to standard Feynman-Schwinger parameter integrals, they are superior to those
in that they already include all possible orderings of the photons, a big advantage in multiloop applications (see below). 

Moreover, the importance of this master formula goes far beyond scalar QED. As discovered by Bern and Kosower, it contains also the full information on the one-loop $N$-photon amplitudes in 
spinor QED, and even on the one-loop (on-shell) $N$-gluon amplitudes in QCD with a scalar, spinor or gluon loop. To extract this information from the master formula, one follows
a certain integration-by-parts procedure, that removes all occurrences of second derivatives, $\GBddtij$, and then applies pattern-matching rules \cite{BK2,BK3,ChrisRev,91}.
Strassler later discovered \cite{Strassler:1992nc} that the same algorithm can be used to derive gauge-invariant decompositions of photon/gluon amplitudes without solving the Ward-identities. 

\subsection{Heat-kernel expansion of the effective action}

Similarly to the propagator case, the effective action can be computed for a general background field by combining the expansion of the interaction exponential with a Taylor expansion
of the field, where the loop center of mass $x_0$ offers itself as the natural expansion point. Performing the path integral over the relative coordinate $q(\tau)$ one obtains the effective
action in the form $\Gamma[A] = \int d^4 x_0 \, {\cal L}(x_0)$, i.e., the point $x_0$ is the one where the effective Lagrangian is evaluated. Again there are various ways of organising the terms of the series, the most 
natural being by mass dimension, which leads to the heat-kernel expansion on the diagonal. However,  using the Green's function $G_B$ the result is not quite identical with the standard heat-kernel
expansion; rather it differs by total derivative terms \cite{EffAcc1}. Those turn out to lead to a somewhat more compact form of the result at high order \cite{EffAcc1,EffAcc2}.  
The standard diagonal heat kernel {\sl would} be obtained, though, if one were to use the Green's function $\Delta$ instead of $G_B$ \cite{EffAcc1}.

Further improvement can be achieved by the choice of Fock-Schwinger gauge centered at $x_0$ \cite{BG0}. In this gauge the coefficients of the Taylor expansion are already written in terms of the field strength tensor:
\bear
A_{\mu}(x) = -\frac{1}{2} F_{\mu\nu}(x-x_0)^{\nu} - \frac{1}{3} \partial_{\lambda}F_{\mu\nu}(x-x_0)^{\nu} (x-x_0)^{\lambda} + \cdots 
\label{FS}
\ear
In this way the expansion becomes manifestly gauge invariant term-by-term. For the simpler case of a scalar self-interaction potential, this method of calculating the heat kernel has been extended to manifolds with boundary 
\cite{Bound2, Bound3} as well as generalised to Robin \cite{BoundR} and semi-transparent boundary conditions see \cite{BoundST}.

\section{Spinor QED: closed loop}

In the case of spinor QED, the closed-loop case is much simpler than the open-line one, so let us start with the former. 

\subsection{Implementation of spin}

The worldline formalism treats spin-half particles not in the usual first-order Dirac formalism, but in a second-order
formalism \cite{SO1, SO2} based on squaring the Dirac operator and using the well-known identity

\bear
(\slash \partial + ie\slash A)^2 =
- (\partial_\mu 
+ i e A_\mu)^2 
-\frac{i}{4}\, e
[\gamma^{\mu},\gamma^{\nu}]
F_{\mu\nu}
\, .
\label{1to2order}
\ear
On the right-hand side we have the Klein-Gordon operator of scalar QED and
a term representing the coupling of the electron spin to the background field. As a consequence, the generalisation of the formula
\eqref{GamScalarPI} to the spinor QED case requires, apart from a global factor coming from statistics and degrees of freedom, 
only the insertion of the following {\it Feynman spin-factor} $\mathcal{S}[x, A]$ under the path integral \cite{Feynman:1951gn}

\bear
\mathcal{S}[x,A] &=& {\rm tr}_{\gamma} {\cal P}
\exp\biggl[{i\frac{e}{4}\,[\gamma^{\mu},\gamma^{\nu}]
\int_0^Td\tau \, F_{\mu\nu}(x(\tau))}\biggr]
\label{defspinfactor}
\ear
where ${\rm tr}_{\gamma}$ denotes the Dirac trace, and
${\cal P}$ is the path ordering operator. 

Thus in the worldline formalism generally any calculation of a quantity in spinor QED involves computing the corresponding quantity in scalar QED, plus additional terms
involving the spin coupling. How to deal with the spin-factor depends on the purpose, though. In a numerical or semiclassical approach it will usually be used as it stands in \eqref{defspinfactor}.
This is not the case in the ``string-inspired'' approach,  where one finds it advantageous to rewrite the spin factor in terms of the following auxiliary path integral
over a Grassmann vector field $\psi^{\mu}(\tau)$ living on the particle's worldline \cite{Fradkin:1966zz,Grass1, Grass2, BdVH,Fradkin:1991ci},

\begin{equation}
	\mathcal{S}[x, A] = \oint \mathscr{D} \psi\, \e^{-\int_{0}^{T} d\tau\, [\frac{1}{2}\psi(\tau) \cdot \dot{\psi}(\tau) - ie \psi(\tau) \cdot  F(x(\tau)) \cdot \psi(\tau)]},
	\label{spinFacPI}
\end{equation}
where the path integral has anti-periodic boundary conditions and normalisation $\oint \mathscr{D} \psi\, \e^{-\int_{0}^{T} d\tau\, \frac{1}{2}\psi \cdot \dot{\psi}} = 2^{\frac{D}{2}}$. 
This allows the incorporation of the spin degrees of freedom into a worldline action 

\begin{equation}
	S[x, \psi | A] = \int_{0}^{T}d\tau\left[ \frac{\dot{x}^{2}}{4} + \frac{1}{2}\psi \cdot \dot{\psi} + ie\dot{x} \cdot A(x(\tau)) - ie \psi^{\mu} F_{\mu\nu}(x(\tau))\psi^{\nu} \right].
	\label{Swlspin}
\end{equation}
The spin path integral will eventually be evaluated using the worldline correlator $G_F(\tau_i,\tau_j) = {\rm sign}(\tau_i-\tau_j)$. 

Note the benefit of the path ordering being absorbed by the path integral of the Grassmann fields and the resulting action enjoying a ``worldline supersymmetry'' (along with rigid translational invariance) under the transformations
%\footnote{\label{footSusy}The transformations (\ref{susyglobal}) follow from (global) diffeomorphisms of a superspace $\tau | \theta$, where $\theta$ is a Grassmann coordinate, where we can define a super-field $X^{\mu}(\tau, \theta) = x^{\mu}(\tau) + \sqrt{2}\theta \psi^{\mu}(\tau)$, under $\tau \rightarrow \tau + \sqrt{2}\theta \wp$ and $\theta \rightarrow \theta - \sqrt{2}\wp$. This global supersymmetry can be gauged to form a one-dimensional supergravity with local diffeomorphism and supersymmetry invariance -- see \cite{BdVH, JO2}. The Faddeev-Popov determinant from gauge fixing this symmetry provides the measure $\frac{dT}{T}$ in the proper time integral, which represents an integration over non-equivalent intrinsic metrics on the worldline.} 
generated by a Grassmann parameter $\wp$ 
\begin{equation}
	\delta x^{\mu} = -2\wp \psi^{\mu}; \qquad \delta\psi^{\mu} = \wp\dot{x}^{\mu}.
	\label{susyglobal}
\end{equation}
This supersymmetry has a significant impact on the evaluation of path integrals throughout this formalism. 

\subsection{The $N$ - photon amplitudes}

The one-loop $N$-photon amplitudes are obtained again by specialising to a plane-wave background and expanding. This leads to a photon vertex operator 
whose form is familiar from the spinning string
\begin{equation}
	V_{\gamma}[k,\varepsilon] = \int_{0}^{T}d\tau\left[ \varepsilon \cdot \dot{x} +2i \ep \cdot \psi\, k \cdot \psi\right] \,\e^{i k \cdot x} \, .
\end{equation}
To arrive at a closed-form master formula for arbitrary $N$, one now needs a way to exponentiate the spin part of the vertex operator. 
A convenient method is through the worldline super formalism \cite{ChrisRev}. However, in practice it turns out to be more efficient to
use an aspect of the Bern-Kosower rules discussed above that makes the explicit computation of the Grassmann path integral unnecessary, namely the ``cycle replacement rule:''
Following calculation of the path integral of the \textit{scalar} effective action, integration by parts can remove all second derivatives of the $\GBij$. 
The replacement rule states that all the spin contributions can be incorporated by replacing all ``closed cycles'' of $\GBdij$ as (a proof of this for general $N$ is in \cite{ChrisRev}):

\begin{equation}
\dot G_{Bi_1i_2} 
\dot G_{Bi_2i_3} 
\cdots
\dot G_{Bi_ni_1}
\rightarrow 
\dot G_{Bi_1i_2} 
\dot G_{Bi_2i_3} 
\cdots
\dot G_{Bi_ni_1}
-
G_{Fi_1i_2}
G_{Fi_2i_3}
\cdots
G_{Fi_ni_1}\, .
\nonumber\\
\label{crr}
\end{equation}
For example, this procedure yields integral representations of the one-loop off-shell four-photon amplitudes in scalar and spinor QED
that are manifestly permutation and gauge invariant, involve only six different tensor structures, combine the various photon orderings, and are free of spurious UV divergences \cite{ChrisRev,91}.
For the massless case, the worldline representation of the one-loop photon amplitudes was used in \cite{Badger:2008rn} to show that, for eight or more photons, they can be reduced to scalar box functions. 

\section{The open fermion line}

Finding a computationally efficient worldline representation for the spin-half propagator coupled to a Maxwell background field is a more difficult, but also very interesting issue. 
Various such representations have been constructed \cite{Fradkin:1991ci,Gitman:1996wm,McKeon:1993sh,Karanikas:2002sy,VanHolten:1995ds,Alexandrou:1998ia} but their practical usefulness remains largely untested. 
A new run on this long-standing issue is presented in \cite{FermProp1} based on a {\it symbol map} representation of the open-line spin factor ($\alpha$ and $\beta$ are spin indices):

\begin{equation}
 \hspace{-1em}\mathcal{S}[x, A]_{\alpha \beta} \equiv	\mathscr{P}\left\{ \e^{-\frac{ie}{4}\int_{0}^{T} d\tau [\gamma^{\mu}, \gamma^{\nu}]F_{\mu\nu}(x(\tau))} \right\}_{\alpha \beta} = 2^{-\frac{D}{2}} \textrm{symb}^{-1}\bigg\{ \int_{\xi(0) + \xi(T) = 2\eta}\hspace{-2em}\mathscr{D} \xi \, \e^{-\int_{0}^{T} \left[ \frac{1}{2}\xi \cdot \dot{\xi} - ie\xi^{\mu}F_{\mu\nu}(x(\tau))\xi^{\nu} \right] -\frac{1}{2} \xi(0)\cdot \xi(T)}\bigg\}_{\alpha \beta},
 \label{spinFacPIS}
\end{equation}
where $\eta^{\mu}$ are constant Grassmann parameters that generate the Dirac matrix structure via the {\it symbol map}

\begin{equation}
	\textrm{symb}\left\{\gamma^{[\mu \nu \ldots \rho]}\right\} \equiv (-i\sqrt{2} \eta^{\mu})(-i\sqrt{2} \eta^{\nu})\cdots (-i\sqrt{2} \eta^{\rho})
\end{equation}
with the square brackets indicating the anti-symmetrised product. By changing variables $\xi^{\mu}(\tau) = \eta^{\mu} + \psi^{\mu}(\tau)$ the boundary conditions become anti-periodic on $\psi^{\mu}$ and the full worldline action for spinor QED becomes

\begin{equation}
	S[x, \psi | A] = \int_{0}^{T}d\tau\left[ \frac{\dot{x}^{2}}{4} + \frac{1}{2}\psi \cdot \dot{\psi} + ie\dot{x} \cdot A(x(\tau)) - ie (\psi + \eta)^{\mu} F_{\mu\nu}(x(\tau))(\psi + \eta)^{\nu} \right].
	\label{SwlspinS}
\end{equation}
In \cite{FermProp1} this approach is shown to be an efficient alternative to Feynman diagrams for calculations such as linear and non-linear Compton scattering and the electron self energy and its tensor decomposition.
For a similar but distinct approach to the open fermion line see \cite{Bhattacharya:2017wlw}.

\section{QED in a constant external field}
Constant external fields play a special role, both for their physical importance and for being one of the few configurations for which the Dirac equation can be solved
analytically, making exact non-perturbative studies feasible (see \cite{HolgerBook} and refs. therein). 
In the worldline formalism, such a background $\hat A_{\mu}$ is incorporated for the closed loop, say, by taking it in Fock-Schwinger gauge \eqref{FS} centered at $x_0$, reading $
\hat A_{\mu}(x) = -\frac{1}{2}\hat{F}_{\mu\nu}{(x - x_{0})}^{\nu}$ so that, from \eqref{Swlspin}, its effect can be absorbed into the kinetic terms of the worldline Lagrangian:
\begin{equation}
	S[x, \psi | \hat{F}, A] = \int_{0}^{T}d\tau\left[ x\cdot \left(-\frac{1}{4}\frac{d^{2}}{d\tau^{2}} +2ie\hat{F}\right) \cdot x + \psi \cdot \left(\frac{1}{2}\frac{d}{d\tau} -ie\hat{F}\right)\cdot \psi + ie\dot{x} \cdot A(x(\tau)) - ie \psi \cdot F (x(\tau))\cdot \psi  \right].
	\label{SwlspinBG}
\end{equation}
This necessitates only a change in the worldline Green's functions, which become \cite{BG1, SP1} ($\Zz \equiv e\hat{F}T$)
\begin{align}
	\GBtij \rightarrow \GBctij = \frac{T}{2\Zz^{2}}\left(\frac{\Zz}{\sin \Zz}e^{-i\Zz \GBdij} + i\Zz \GBdij - 1\right); \quad 	\GFtij \rightarrow \GFctij = \GFij\frac{e^{-i \Zz \GBdij}}{\cos \Zz},
\end{align}
and the free path integral normalisations change to
\begin{align}
	\Det\left[-\frac{1}{4}\frac{d^{2}}{d\tau^{2}} +2ie\hat{F}\right] = (4\pi T)^{-\frac{D}{2}}\det{}^{-\frac{1}{2}}\left[\frac{\sin \Zz}{\Zz} \right]; \qquad \Det\left[\frac{1}{2}\frac{d}{d\tau} -ie\hat{F}\right] = 2^{\frac{D}{2}}\det{}	^{-\frac{1}{2}}\left[\cos \Zz\right].
\end{align}
With $\hat F$ the only field one reproduces the one-loop Euler-Heisenberg Lagrangians \cite{BG0, EH1} as determinants:
\begin{align}
	\mathscr{L}^{(1)}(\hat{F}) &= \int_{0}^{\infty}\frac{dT}{T}(4\pi T)^{-\frac{D}{2}}\e^{-m^{2} T} \det{}^{-\frac{1}{2}}\left[\frac{\sin \Zz}{\Zz} \right]\, ; \qquad \quad \,\,\, \textrm{Scalar QED}, \\
	\mathscr{L}^{(1)}(\hat{F}) &= -2\int_{0}^{\infty}\frac{dT}{T}(4\pi T)^{-\frac{D}{2}}\e^{-m^{2} T} \det{}^{-\frac{1}{2}}\left[\frac{\tan \Zz}{\Zz} \right]\, ; \qquad \textrm{Spinor QED}.
\end{align}
To obtain the one-loop $N$-photon amplitudes in the constant field background one may re-use the Master Formulae (\ref{GamScalar}) with the replacement of the Green's functions and functional determinants by their constant field counterparts  \cite{BG1,SP1}.
Furthermore, the integration-by-parts procedure and cycle replacement rules carry forward with minor modifications \cite{ChrisRev}.
In the constant background field case, the resulting formalism is generally much superior to the Feynman diagrammatic one,
since treating the loop as a whole,
rather than segmenting it into individual propagators, becomes more technically advantageous
in the presence of the background field where those propagators themselves already possess a complicated Lorentz structure. 
This has been demonstrated in a wide number of applications. For the closed-loop case, those include 
the scalar/spinor QED vacuum polarisation tensors in a general constant field \cite{SODit, VacPol1},
photon splitting in a magnetic field \cite{AdChris}, and the heat-kernel expansion modified by the constant background in four \cite{EA1} 
and in three dimensions \cite{EffAcc3}. For a recent application to Schwinger pair creation see \cite{Torgrimsson:2017pzs}. 

The open-line case had been treated for scalar QED for a purely magnetic field in \cite{mckshe}, which was later generalised to the general
constant field case in \cite{lineBG}. For the open fermion line, as in the vacuum case until recently little had been done. However, this is now changing; in \cite{FermProp2}
it will be shown how to extend the approach of \cite{FermProp1} mentioned above to include the constant field background,  
and a corresponding master formula will be given. See also \cite{GKscalar,GKspinor} where the reducible one-loop contributions to the scalar resp. fermion
propagator in a constant field are computed along these lines. 

\section{Multi-loop amplitudes}

Our parameter integrals and Master Formulae are valid off-shell, so that they can be used to construct multi-loop amplitudes via the sewing of pairs of external legs. 
Alternatively, as in string theory, one can incorporate internal propagators into the worldline Green's functions, leading to the concept of ``Green's functions on graphs''  \cite{multi1,Sato:1995xc,Roland:1996np,Dai:2006vj}.
This has turned out to be very efficient for calculating two-loop Euler-Heisenberg Lagrangians for various constant field configurations, in scalar as well as in spinor QED \cite{SP1,PC1}. 
The two approaches are equivalent, though, and lead to the same final integral representations. 

\begin{figure}[h]
  \centering
   \includegraphics[width=0.75\textwidth]{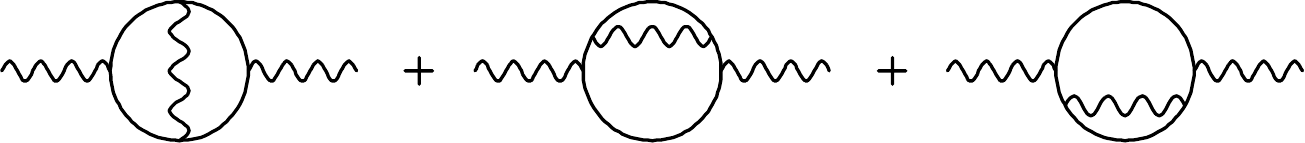}
\caption{Diagrams contributing to the 2-loop QED $\beta$ function.} 
\label{fig-2loopvp}
\end{figure}

One interesting aspect of these representations is that they often allow one to unify 
Feynman diagrams of various topologies since the master formula \eqref{GamScalar} includes all possible orderings of the photon legs. Taking the four-point case and sewing up two of the four photons legs
with a photon propagator, one obtains the diagrams shown in Fig. \ref{fig-2loopvp}, that is the two-loop photon propagator. 

\noindent This sum involves already two different topologies, which in a Feynman
diagram calculation would have to be computed separately. At the same time, all practicians of QED know that precisely this type of sum often suffers from large cancellations 
between diagrams, related to gauge invariance \cite{Cvitanovic:1977dp}. In \cite{multi2} it was shown how to use the worldline formalism for computing the UV-divergent part of these diagrams (the QED $\beta$-function)
dealing with the three diagrams as a a whole, and without encountering any non-trivial integrals. 

\begin{figure}[h]
  \centering
   \includegraphics[width=0.4\textwidth]{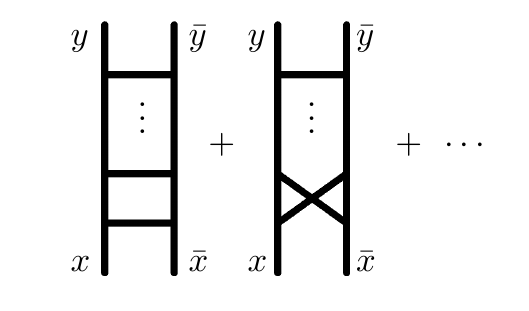}
\caption{Ladder and crossed-ladder diagrams in scalar field theory.}
\label{fig-ladders}
\end{figure}

A multi-loop application to scalar field theory was given in \cite{Bastianelli:2014bfa}, using the sewing procedure to obtain an integral representation that
unifies the ladder and crossed-ladder diagrams shown in Fig. \ref{fig-ladders}. 	

\noindent This integral representation was used with various extrapolations and saddle-point approximations to estimate the lowest bound-state energy
in the case of a massless scalar exchange particle (rungs of the ladders).

\section{Non-abelian gauge theory}

Next we extend the string-inspired formalism to non-abelian gauge theory. As mentioned, the seminal work of \cite{BK2,BK3,Strass1} focussed on one-loop gluon amplitudes. 
There are two ways of dealing with the additional colour degree of freedom, one straightforward, one recent and quite subtle, which we discuss in turn.

\subsection{Explicit colour factors}

%Returning to Feynman's worldline formula for the effective action \eqref{GamScalarPI}, it is easy to guess how to adapt it to the non-abelian case \cite{Strass1}:
It is easy to guess how to adapt Feynman's representation of the effective action, \eqref{GamScalarPI}, to the non-abelian case \cite{Strass1}: the gauge-field must be become a non-abelian one, $A_{\mu} = A_{\mu}^aT^a$ with $T^a$ the generators of the colour group in the representation of the loop scalar. This will in general mean that the worldline
Lagrangian at different proper-times does not commute with itself any more, so the exponential is replaced by a path-ordered one. Finally, a global colour trace makes the
effective action a scalar. 

The $N$-gluon amplitudes can then be obtained following the same procedure as above for the $N$-photon amplitudes, leading to a gluon vertex-operator that differs from \eqref{defvertop} only
by an additional colour matrix factor $T^a$. The resulting master formula for the $N$-gluon amplitudes differs from \eqref{GamScalar} only by a global colour factor ${\rm tr}(T^{a_{i_1}}\cdots T^{a_{i_N}})$
and by the parameter integrals $\int d\tau_i$ being ordered, i.e. calculated for a fixed cyclic ordering of the gluons along the loop. 
In QCD one is more interested in the fermion and gluon loop contributions to the gluon amplitudes, but unlike the Feynman diagram approach these are not separate calculations here, rather
the passage from the scalar to the spinor and gluon loop cases can be effected by an extension of the cycle-replacement rule \eqref{crr} above \cite{BK2,BK3,Strass1,SP1}.
Differently from the abelian case, here there are also reducible contributions, where not all the
gluons sit directly on the loop; as one of the remarkable properties of the Bern-Kosower formalism, the information on those can also be retrieved from the master formula. 
This technology was used in \cite{Bern:1993mq} for the first computation of the one-loop five-gluon amplitudes. 

While the work of Bern and Kosower concentrated on the on-shell case, the worldline approach works off-shell too; in particular, the ``cycle-replacement rules'' have been rederived
inside the worldline formalism \cite{Strass1,SP1}. Together with the already mentioned fact that the formalism allows one to generate gauge-invariant structures by certain integration-by-parts algorithms, 
which becomes much more non-trivial in the non-abelian case \cite{Strassler:1992nc}, this has led to the recognition that the formalism is ideally suited to form-factor decompositions
of the $N$-gluon amplitudes; see \cite{91,Ahmadiniaz:2016qwn} for more details. 

Also the the heat-kernel expansion can be extended to the non-abelian case without difficulty, 
including the use of Fock-Schwinger gauge. In particular, in \cite{Fliegner:1997rk} it was employed for computing the non-abelian effective action to fifth order in the heat-kernel expansion. 

\subsection{Auxiliary worldline colour fields}
For spinor QED, a main motivation for the replacement of the Feynman spin factor \eqref{defspinfactor} by a Grassmann path integral in \eqref{spinFacPI}
comes from the desire to be able to work without having to fix the ordering of the photons. This suggests attempting a similar strategy in the presence of colour, i.e.
to replace explicit colour matrices by auxiliary path integrals as we now describe.

That gauge group degrees of freedom can be described by the introduction of auxiliary worldline fields is quite an old observation \cite{Barduc1, Barduc2}. 
These ``colour fields'' can be anti-commuting (Grassmann valued) as in \cite{Col1} or bosonic variables \cite{Col2}. 
With $F$ copies of auxiliary colour fields, $\bar{c}_{f}$ and $c_{f}$, the spin $1/2$ phase space action takes the locally supersymmetric form \cite{JO1, CreteProc}
\begin{equation}
	S[p, x, \psi, e, \chi, \bar{c}, c | A] = \int_{0}^{1}d\tau \left[p \cdot \dot{x} + \frac{i}{2}\psi \cdot \dot{\psi} + \sum_{f = 1}^{F} i \bar{c}^{r}_{f} \dot{c}_{rf} - e\widetilde{H} -i\chi \widetilde{Q} - \sum_{f=1}^{F}a_{f}(L_{f} - s_{f}) - \sum_{g < f}a_{fg}L_{fg}\right].
	\label{Scol}
\end{equation} 
The colour fields have canonical Poisson brackets $\{\bar{c}^{r}_{f}, c_{gs}\}_{\pb} = (-1)^{F}i\delta_{fg}\delta^{r}_{s}$ and the indices $r$ and $s$ indicate they transform in the (conjugate-)fundamental representation of $SU(N)$. The colour fields provide a classical representation of the gauge group algebra by defining $R^{a}_{f} = \bar{c}^{r}_{f}(T^{a})_{r}{}^{s}c_{fs}$ which satisfy $\{R^{a}_{f}, R^{b}_{f}\}_{\pb} = f^{abc}R^{c}_{f}$, where the  $f^{abc}$ are the structure constants. This has been used in (\ref{Scol}) to absorb the gauge group indices in the Hamiltonian, $\widetilde{H} = \widetilde{\pi}^{2} + \frac{i}{2}\psi^{\mu}F^{a}_{\mu\nu}\psi^{\nu} \sum_{f = 1}^{F}R^{a}_{f}$ and  supercharge $\widetilde{Q} = \psi \cdot \widetilde{\pi}$, where $\widetilde{\pi}^{\mu} = p^{\mu} - \sum_{f =1}^{N}A^{a\mu}R^{a}_{f}$. The path ordering prescription is yielded automatically upon integration over $\bar{c}$ and $c$ and the full field strength tensor enters $\widetilde{H}$ whilst preserving worldline supersymmetry.

In (\ref{Scol}) the worldline gauge fields $a_{fg}(\tau)$ partially gauge the $U(F)$ symmetry rotating the colour fields generated by occupation number operators $L_{fg} \equiv \bar{c}^{r}_{f}c_{rg}$, allowing for the Chern-Simons terms associated to the diagonal elements as in (\ref{SHS}) below. The colour fields' Hilbert space is described by (the reducible) $F$-fold tensor products of all possible fully (anti-)symmetric representations of the gauge group. In path integral quantisation on the loop, gauge fixing the $U(F)$ fields to $a_{fg} = \textrm{diag}(\theta_{1}, \ldots, \theta_{F})$ leads to a Faddeev-Popov determinant that acts as a measure on the constant angular moduli. Integration over the moduli projects onto an irreducible subspace of the Hilbert space whose symmetries are fixed by the integers $s_{f}$ in (\ref{Scol}).

In \cite{JO1} it was verified that this produces the correct number of degrees of freedom for an $SU(N)$ multiplet transforming in an arbitrarily chosen irreducible representation, using either Grassmann or bosonic colour fields. Including the gauge field interaction generates the full non-Abelian Wilson-loop interaction of the chosen multiplet \cite{JO2}. This was utilised in \cite{Col1, Col2} for the spinor and scalar loop contributions to gluon vacuum polarisation and to derive a parameter integral expression for $N$-point one-loop gluon amplitudes. Lately this has also been applied to the scalar propagator and tree-level gluon amplitudes \cite{ColTree}. 
A unified theory based upon the standard model, the anti-symmetric multiplets of (flipped-) $SU(5)$, and representations of $SO(10)$ and $SO(16)$ can be produced from $F=1$ family of colour fields without a projector \cite{Paul, MeUnif}. Colour fields are also fully compatible with the new representation of the open fermion line presented in \cite{FermProp1}.

\section{Yukawa and axial couplings}
\label{secYuk}
To determine arbitrary standard model amplitudes one also needs the coupling of fermions to scalars, pseudoscalars and axial vectors. Suitable worldline Lagrangians were derived in the nineties
along various routes \cite{Yuk, Ax, YukAx1, YukAx2, McK1, McK2}. The scalar plus pseudo-scalar case can be treated elegantly by dimensional reduction from $D=6$ \cite{Yuk}, 
setting $A_{M} = (A_{\mu}, \phi_{5}(x^{\mu}) , \phi(x^{\mu}) + m_{\phi})$ where $\mu \in \{1, \ldots 4\}$ and $m_{\phi}$ will be the fermion mass as in the Higgs mechanism. For vanishing $A_{\mu}$ the Gaussian path integral over $x^{5}$ and $x^{6}$ appends to the four dimensional worldline action  (we set the coupling constant $e \rightarrow g$ for this section)

\begin{equation}
	\delta S[x, \psi | A_{\mu} = 0] = \int_{0}^{1} d\tau \left[ \frac{1}{2}\psi_{5}\cdot \dot{\psi}_{5} + \frac{1}{2}\psi_{6}\cdot \dot{\psi}_{6} + g^{2}\phi^{2} + + 2mg\phi + g^{2}\phi_{5}^{2} + 2ig\psi_{5} \psi \cdot  \partial \phi_{5} + 2ig \psi_{6} \psi \cdot   \partial \phi \right].
\end{equation}
Insertion of plane waves for the (pseudo-)scalars, the path integral quantisation of the full action leads \cite{ChrisRev} to a Master Formula for the real part of the $N$-point one-loop (pseudo)-scalar amplitudes with Yukawa coupling to the fermion loop. Contrasting with the standard formalism,  it is written already in terms of scalar integrals, so that the usual tensor reduction is not necessary. 
This observation led to the calculation of the one-loop six-point Yukawa amplitude in \cite{48} (for the massless loop), which was the first computation of a one-loop six-point amplitude involving a fermion loop. 
However, dimensional reduction does not yield the imaginary part of the effective action or amplitude, which also exists in the presence of pseudo-scalars (except for the massless case). 
For incorporation of the imaginary part see \cite{YukAx1,YukAx2}. 

The axial vector case has been treated in \cite{Ax, YukAx1,YukAx2} separating into real and imaginary parts, but there is also a treatment that avoids this separation \cite{McK1, McK2}, at the cost of
working with a non-hermitian exponent. It reproduces the chiral anomaly \cite{McK1} with the anomalous divergence fixed to be in the axial vector current. 
It can be generalised to include an additional constant external (vector) field, which was used in \cite{VacPol2,VacPol3} to determine the vector-axialvector and axialvector-axialvector 
two-point functions in a constant field.

The various approaches have in common that the $\gamma_5$ matrix is represented by the fermion parity operator, $(-1)^{F}$, which after expansion of the path integral sets the boundary conditions on the Grassmann fields, $\psi^{\mu}$, to be anti-periodic (periodic) according to whether it is raised to an even (odd) power. For periodic boundary conditions, a zero mode appears in the spectrum of the first derivative operator, and its saturation
%and it is the saturation of this zero mode that
 in $D=4$ leads to the appearance of the antisymmetric tensor $\varepsilon_{\alpha\beta\gamma\delta}$, just like in string theory. 

\section{Gravitational interactions - QFT in curved space}

Including gravity requires the construction of the worldline path integral in curved space, leading to difficulties with the preservation of general covariance that have been known in quantum
mechanics for decades, but were fully resolved only in the framework of the worldline formalism (see \cite{BastBook} and refs therein). 
Analogous problems exist already with gauge invariance in flat space, but here they can be solved simply and completely defining
the path integral by the usual time-slicing discretisation procedure and the mid-point rule (corresponding to the Weyl ordered Hamiltonian) that preserves gauge invariance \cite{Reg1, Reg2}.
Applying the same procedure to a scalar field coupled to a gravitational background leads to a worldline action

\begin{equation}
	S[x | g]_{W} = \int_{0}^{T}d\tau \left[\frac{1}{2} g_{\mu\nu}(x(\tau))\dot{x}^{\mu}\dot{x^{\nu}} + \frac{\hbar^{2}}{8}\left(R(x(\tau)) + g^{\mu\nu}(x(\tau))\Gamma^{\rho}_{\mu\sigma}(x(\tau))\Gamma^{\sigma}_{\nu\rho}(x(\tau))\right)\right]
\end{equation}
where the order $\hbar^{2}$ terms are a remnant of the Weyl ordering of the Hamiltonian \cite{GravTS}. 
It turns out, however, that alternative regularisations of the path integral affect these $\hbar^{2}$ terms. Mode regularisation \cite{mode1, mode3}, for example, changes the term involving the Christoffel symbols, 
whilst dimensional regularisation \cite{dimrg1, dimrg2} removes these non-covariant terms, leaving only the piece involving the Ricci scalar. 
One should also include ghost degrees of freedom \cite{GravGh1, GravGh2} to take into account the path integral measure in curved space
\begin{equation}
\int	\mathscr{D}_{g}x(\tau) \sim \int \lim_{N\rightarrow \infty} \prod_{i = 1}^{N}\sqrt{|g(x(\tau_{i}))|}d^{D}x(\tau_{i}) = \int \mathscr{D}x(\tau)\mathscr{D}b(\tau) \mathscr{D}c(\tau) \, \e^{-\int_{0}^{T}d\tau b(\tau) \sqrt{|g(x(\tau))|}  c(\tau)}
\end{equation}
where the ghosts have weight $1/2$ and the path integral measures on the right hand side are translationally invariant.
For the explicit computation of the path integral it is often convenient to use Riemann normal coordinates to simplify the expansion of the metric, which is analogous to the use of Fock-Schwinger gauge in gauge theory. 

Applications include the determination of trace anomalies (see \cite{BastBook} and refs. therein), 
gravitational corrections to the Euler-Heisenberg Lagrangian \cite{EHLG1}, one-loop photon-graviton mixing in an electromagnetic field \cite{PG25, PG3} and the photon vacuum polarisation tensor in curved space \cite{Hollowood:2007ku}. 
The worldline action also generalises to the fermionic case maintaining an $N=1$ supersymmetry as in flat space \cite{GravSP1, GravSp2, GravSp3}. This has been applied to determine the graviton self-energy \cite{GravSE} (see also \cite{GravSEs} for the scalar case) and further trace anomalies \cite{GravTrF}. 
Extension to vector and anti-symmetric tensors and their contribution to the graviton self energy was achieved in \cite{GravVec2, GravVec1} (see also the following section). 
Recent work has also revisited trace anomalies on maximally symmetric spaces \cite{Symm1, Symm2}.

\section{Higher spin fields}
\label{secHS}

Interest in higher spin fields \cite{HS1, HS2} has been stimulated by string theory and holography \cite{HS4, HS3}. Models of particles with arbitrary spin were proposed long ago \cite{HSold2, HSold3} and the worldline formalism relates these systems to higher spin fields. The worldline theory is most easily studied in phase space, the archetypal action being the $O(N)$ extended locally supersymmetric point particle \cite{HSold1, HSWL1, HSWL2}
\begin{equation}
	S[p, x, \psi, e, \chi, a] = \int_{0}^{1} d\tau \left[p \cdot \dot{x} + \sum_{i=1}^{N} \frac{i}{2}\psi_{i} \cdot \dot{\psi}_{i} - eH - \sum_{i = 1}^{N}i \chi_{i} Q_{i} - \sum_{i, j = 1}^{N} \frac{1}{2}a_{ij}\left(J_{ij} - n_{i}\delta_{ij}\right)  \right]
	\label{SHS}
\end{equation}
where we have rescaled to the unit circle and returned to Minkowski space. We have introduced an einbein, $e(\tau)$, $N$ (Grassmann) gravitinos $\chi_{i}(\tau)$ and $O(N)$ worldline gauge fields $a_{ij}(\tau)$ to gauge the first class constraints from the Hamiltonian, $H = \frac{p^{2}}{2}$, the supercharges $Q_{i} = p \cdot \psi_{i}$ and $R$-charges $J_{ij} = i\psi_{i} \cdot \psi_{j}$. At each spatial point the particle wavefunction has components transforming in the $s$-fold tensor products of all representations with up to $D$ fully antisymmetric Lorentz indices -- see \cite{HSWL3}. 

In fixing the local supersymmetry, the $a_{ij}$ can be set to be constants that are moduli of the worldline gauge fields leading to finite dimensional integrals over the moduli against their Faddev-Popov determinant. These integrals impose the first class constraints which enforce unitarity and irreducibility, controlled by the Chern-Simons terms in (\ref{SHS}). The resulting wavefunction satisfies the Bargmann-Wigner equations \cite{Barg1} for massless free spin $s = \frac{N}{2}$ fields \cite{Barg2, HSold3}. For the BRST quantisation of such models see \cite{BRST1, BRST2, BRST3}.  The details depend very much on the model and so we briefly recap applications in the literature.

Spin $1$ vector and antisymmetric tensor particle models \cite{AsymT} were used in \cite{ASWL1, ASWL2} to calculate Seeley-DeWitt coefficients of the heat kernel for tensor fields and their contribution to the graviton self energy. Conformal fields were described in \cite{HSWL1} based on an $SO(N)$ version of (\ref{SHS}). The approach was generalised to include coupling to conformally flat metrics, such as an (A)dS background \cite{HSDS1}, to derive the one-loop effective action and heat kernel coefficients \cite{HSDS2, HSC1}. Higher spin fields on complex manifolds require a $U(N)$ invariant form of (\ref{SHS}) which describes massless differential forms on K\"{a}hler manifolds and the short-time heat kernel expansion \cite{HSK1, HSK2, Forms1, Forms2}. A review can be found in \cite{HSWL3} and application to position space transition amplitudes at \cite{HSRev}.

\section{Non-commutative space-time}
\label{secNC}

Field theory on a non-commutative space-time \cite{Madore:2000en,NC1, NC2} has deep connections with string theory and quantum gravity \cite{NCGrav, NCLec}. Its non-locality and minimal length scale leads to UV/IR mixing and other interesting renormalisation properties \cite{NC3} that can be studied using the worldline formalism \cite{NCWL1, NC4, SatoNC}. The specific model studied had non-commutative structure $[x^{\mu}, x^{\nu}] = 2i\Theta^{\mu\nu}$ that is most easily expressed through the Moyal $\star$-product \cite{Moyal1, Moyal2} deforming usual multiplication
\begin{equation}
	\Psi(x)\star \Phi(x) \equiv \Psi(x)\exp\left(\frac{i}{2}\overleftarrow{\partial}_{\mu}\theta^{\mu\nu}\overrightarrow{\partial}_{\nu}\right)\Phi(x) \, .
\end{equation}
The authors of \cite{NCWL1} considered uncharged $\Phi_{\star}^{3}$ and $\Phi_{\star}^{4}$  theory \cite{NCScal}
%\footnote{By which we mean the field theory action for a $\Phi_{\star}^{4}$ interaction is $\int d^{4}x \, \left[\frac{1}{2} \partial_{\mu}\Phi \partial^{\mu}\Phi  + \frac{m^{2}}{2}\Phi^{2} + \frac{\lambda}{4!}\Phi \star \Phi \star \Phi \star \Phi \right]$.}
. Due to non-locality the worldline theory is simplest to study in phase space, where the effective action is produced with worldline action
\begin{equation}
	S_{\Theta}[p, x] = \int_{0}^{T}d\tau \left[p^{2} - i p \cdot \dot{x} + \frac{\lambda}{3}\bigg(\Phi_{\star}^{2}(x + \Theta\cdot p) +  \Phi(x +  \Theta \cdot p)\Phi(x -  \Theta \cdot p) + \Phi_{\star}^{2}(x - \Theta \cdot p)\bigg) \right].
	\label{SNC}
\end{equation}
In \cite{NCWL1} a generating function for the Seeley-deWitt coefficients in the (non-local) heat kernel expansion was provided and the (non-)renormalisability of the field theory was studied.

The Grosse-Wulkenhaar model \cite{GW1, GW2} shows how renormalisability can be restored in scalar field theory by introducing a harmonic oscillator background \cite{GW3}, $\mathcal{L}_{\textrm{HO}} = \frac{\omega^{2}}{2}x^{2}\Phi^{2}$ to the field theory Lagrangian. This model was considered from a worldline perspective in \cite{GWWL}, adding a piece $\omega^{2}x^{2}$ to the integrand of (\ref{SNC}). Of course, being quadratic in the trajectory $x(\tau)$, this can be absorbed into the kinetic term of the point particle mode. The one-loop effective action of this model was determined and the two- and four-point functions examined. Later the same formalism was applied \cite{NCWLU} to $U(1)$ Yang-Mills theory \cite{NCU1, NCU2}. This $U(1)^{\star}$ model has the phase space worldline action (no auxiliary fields were used in \cite{NCWLU}) 
\begin{equation}
	S_{\gamma \star}[p, x | A] = \int_{0}^{\infty}d\tau \left[ -i p \cdot \dot{x} + \delta_{\mu\nu}\big(p + A(x + \Theta \cdot p) - A(x - \Theta \cdot p) \big)^{2} - 2i\big(F_{\mu\nu}(x + \Theta \cdot p ) - F_{\mu\nu}(x - \Theta \cdot p)\big)\right]
\end{equation}
This model was analysed in \cite{NCWLU} (for a related approach see \cite{Kiem}) to find the two- and four-point functions and the $\beta$-function of $U(1)^{\star}$ theory \cite{UVIR}; note again that the $p^{2}$ term from the  first brackets can be absorbed into the free particle action for each component of the matrix valued path integral  whose trace is taken at the end. Ongoing work is generalising this to the $U(N)^{\star}$ with the use of auxiliary fields \cite{JOPN}.

\medskip

\noindent{\bf Acknowledgements:}
The authors are very grateful to Petr Jizba for the invitation to attend the Path Integrals in Complex Dynamical Systems workshop and for the opportunity to contribute this technical report, and to the workshop organisers, Petr Jizba, Stefan Kirchner, Lawrence Schulman and Jan Zaanen. CONACYT provided financial support through project CB2014-242461.

\bibliographystyle{elsarticle-num}
\bibliography{bibRepchris}

\end{document}